\documentstyle[12pt]{article}

\def\cD{{\cal D}}
\def\a{\alpha}

\def\d{\delta}
\def\l{\lambda}
\def\lb{\bar{\lambda}}
\def\s{\sigma}
\def\sb{\bar{\sigma}}
\def\th{\theta}
\def\thb{\bar{\theta}}
\def\psib{\bar{\psi}}
\def\xib{\bar{\xi}}
\def\etab{\bar{\eta}}
\def\Db{\bar{D}}
\def\dg{\dagger}
\def\half{\frac{1}{2}}

\newskip\humongous \humongous=0pt plus 1000pt minus 1000pt

\newif\ifdtup

\def\beq{\begin{equation}}
\def\eeq{\end{equation}}

\def\beqn{\begin{eqnarray}}
\def\eeqn{\end{eqnarray}}
\relax

\def\G2{{\; \rm GeV/}c^2}
\def\G{\; \rm GeV}

\def\dotx{\dotx{\dot\overline{x}}}

\relax

\def\p{\partial}

\begin{document}
\begin{titlepage}
\begin{flushright}
       {\normalsize  OU-HET 276 \\  hep-th/9708123\\
             August, 1997,  October revised }
\end{flushright}
%
\begin{center} 
  {\large \bf $USp(2k)$ Matrix Model :  $F$ Theory Connection
 }\footnote{This work is supported in part 
 by the Grant-in-Aid  for Scientific Research Fund (2126)
from the Ministry of Education, Science and Culture, Japan.}

\vfill
         {\bf H.~Itoyama}  \\
            and \\
         {\bf A.~Tokura}\\
        Department of Physics,\\
        Graduate School of Science, Osaka University,\\
        Toyonaka, Osaka, 560 Japan\\
\end{center}
\vfill
\begin{abstract}
   We present a zero dimensional matrix model based on $USp(2k)$
 with supermultiplets in symmetric, antisymmetric and fundamental
 representations.  The four dimensional compactification of this model
 naturally captures the exact results of Sen \cite{Sen} in $F$ theory.
 Eight dynamical and eight kinematical supercharges are
 found, which is required for critical
 string interpretation. Classical vacuum has ten coordinates  and is
 equipped with orbifold structure.
 We clarify the issue of spacetime dimensions  which $F$ theory  represented
 by this matrix  model produces.
\end{abstract}
\vfill
\end{titlepage}

\section{Introduction}

  Recently  there are a few interesting directions  emerging on the
 formulation  of strings. On the one hand, nonperturbative formulation of
 string theory as matrix models
  via the notion of noncommuting coordinate \cite{Witten} is developing.
   This includes  vigorous activities on the large $N$  quantum mechanical
 model \cite{BFSS} which formulates M theory \cite{M} as well as the
 zero dimensional model \cite{KEK,2B} of type $IIB$ superstrings.  Notion of
 string compactification and attendant
 counting of degrees of freedom appear to be very different from what 
 we thought of unification based on perturbative strings. So far we have been
  able to discuss only toroidal compactifications with/without
  discrete projection \cite{BFSS,comp,het,WT} through a specific
 procedure \cite{WT}.

 The other aspect includes the developments centered
 around $F$ theory \cite{V}. This provides a new perspective to
  treating type IIB strings on exact quantum backgrounds
  through purely geometrical framework.
 $F$ theory captures an intriguing phenomenon of 
  string coupling depending on internal space\footnote{ In the conventional
 approach  of the first quantized strings, this is  physics related to
 the orientifold compactification\cite{PW,AIKT} } beyond perturbative
  consideration.
   One way of viewing this $F$ theory is that it provides with us a new scheme
 of compactifications of string theory  which are defined beyond
  perturbation theory.  This is the point of view we wish to adopt in the
 present paper.  A series of compactifications whose pertubative limit  are
those of orientifold are prototypical examples.

  In this paper,  we wish to give this scheme of $F$ theory  a
  constructive framework as a matrix  model. 
 We present a $USp(2k)$ matrix model in zero dimension
  and discuss several properties. We argue  that our model in the particular
  large $N$ limit  produces an exact $F$ theory compactification. 
   The model consists of matrices belonging to the symmetric (adjoint)
 and antisymmetric representations, and $n_{f}$ of  $2k$
 dimensional vectors.
 The $n_{f}=4$ and $n_{f}=16$ cases stand out of special significance. 
 The large $k$ limit will capture string physics in the
 sense of t'Hooft.  The model is inspired in the $n_{f}=4$  case
 by the supermultiplets of the UV finite ${\cal N} =2$ 
 supersymmetric gauge theory in four dimensions with gauge group $USp(2k)$.
  We are motivated by the exact result of Sen \cite{Sen} in $F$ theory in
 eight spacetime dimensions on certain elliptic fibered $K3$.
 The exact description of axion/dilaton sector on this $F$ theory
 compactification has been found to be mathematically identical to that of the
 quantum moduli space \cite{SW} of the susy gauge theory above.  ( See
\cite{BDS,DLSASYT}.)
  We thus see that our model ( the $n_{f}=4$ case ) after four dimensional
 compactification under the procedure of \cite{WT}  possesses the above
 quantum moduli space and naturally reproduces Sen's result in $F$ theory
  in eight spacetime dimensions.   We will discuss this again in the end of
 this paper.

 In the next section, we define the $USp(2k)$ matrix model in zero dimension.
  Eight dynamical and eight kinematical supercharges
  are shown to exist in our model in section three.   This is necessary for
  this model to be interpreted as  critical string theory ${\it i.e.}$
    unified   theory of gravity and other forces.

 In section
 four, we determine the classical vacuum,  which is found to be labelled by
  ten coordinates.  The one-loop stability of this
 geometry is  ensured by supersymmetry.  
   We discuss  the case in which  the model in the particular large $k$ limit
  produces  an exact $F$ theory compactification.
   This is the compatification whose  perturbative  limit
  is described by the eight dimensional type $IIB$ string on $T^{2}/Z^{2}$
  orientifold. The issue of  twelve versus ten spacetime dimensions
 naturally emerges.   We clarify the sense of the spacetime dimensions
 designated by the model in the particular large $k$ limit (compactification).

 We adopt a notation that the inner product of  two $2k$ dimensional
 vectors  $u_{i}$ and $v_{i}$ invariant under $USp(2k)$  are
\beqn
 \langle u, v \rangle = u_{i} F^{ij} v_{j} \;, \;\;  {\rm with} \;\; 
  F^{ij} =  \left( \begin{array}{cc}
                  0 & I_k \\
                  -I_k & 0
                  \end{array}
            \right) \;\;.      
\eeqn
 Here $I_k$ is the unit matrix.  We may regard $\left( u_{i} \right)^{*}
\equiv \left( u^{*}\right)^{i}$.
 Raising and lowering of the indices are  done  by $F=F^{ij}$
  and $F^{-1} = F_{ij}$.
   Any element $X$ of the $usp(2k)$ Lie algebra
  satisfying  $X^{t}F + FX =0$ and $X^{\dagger} = X$ can be represented as
\beqn
\label{eq:usp}
  X =\left(
     \begin{array}{cc}
           M     &     N    \\
           N^{*} &  -M^{t}  
     \end{array}
            \right)  \;, \;\;
 {\rm with}\;\;  M^{\dagger} =M \;, \;\; N^{t} =N\;.
\eeqn
  Chiral superfields are expanded by the  generators of $usp(2k)$
 with coefficients being complex.


\section{Definition of the zero dimensional matrix model}

    Our zero-dimensional model can be written, by borrowing
 $N=1$, $d=4$ superfield notation in the Wess-Zumino gauge. One  simply drops
   all spacetime dependence of the fields but keeps all grassmann coordinates
  as they are
\beqn
   S &\equiv&  S_{ {\rm vec}} + S_{{\rm asym}} + S_{{\rm fund}} \;\; 
\\
  S_{ {\rm vec}} &=& \frac{1}{4 g^{2}} \;  Tr \left(
 \int d^2 \th W^{\a} W_{\a} + h.c. +
4 \int d^2 \th d^2 \thb \Phi^{\dg} e^{2V} \Phi e^{-2V} \right)
 \nonumber   \\
  S_{{\rm asym}} &=&    \frac{1}{g^{2}} \int d\theta^{2} d \bar{\theta}^{2}
 \left( T^{* \; ij} \left( e^{2V({\rm asym} ) }
 \right)_{ij}^{\;\;\;k \ell}
  T_{k \ell} +  \tilde{T}^{ij} \left( e^{-2V({\rm asym} ) }
 \right)_{ij}^{\;\;\;k \ell}
 \tilde{T}^{* }_{k \ell} \right)
\nonumber\\
& & + \frac{1}{g^{2}} \left\{ \sqrt{2} \int d\theta^{2}
\tilde{T}^{ij} \left( \Phi_{(asym)}
 \right)_{ij}^{\;\;\;k \ell}
T_{k \ell}   + h.c.
 \right\}
\nonumber
\\
  S_{{\rm fund}} &=&  \frac{1}{g^{2}} \sum_{f=1}^{n_f}
\left[
  \int d^2 \th d^2 \thb
\left( Q_{(f)}^{* \;i} \left( e^{2V} \right)_i^{\;\;j} Q_{(f) \; j}
+  \tilde{Q}_{(f)}^{ i} \left( e^{-2V} \right)_i^{\;\;j}
        \tilde{Q}_{(f) \; j}^{*} \right)
\right.
\nonumber
\\
& &
\qquad
 +
\left.
\left\{
\int d^2 \th
\left(
	m_{(f)} \tilde{Q}_{(f)}^{\;\;\;\; i}  Q_{(f) \;i}
	+ \sqrt{2} \tilde{Q}_{(f)}^{\;\;\;\; i}
		 \left( \Phi \right)_i^{\;\;j} Q_{(f) \;j}
\right)
+ h.c.
\right\}
\right]  \;\;\;.
   \nonumber
\eeqn

  The chiral superfields introduced above  are
\beqn
    W_{\alpha} &=&  -\frac{1}{8} \Db \Db e^{-2V} D_{\a} e^{2V}
\;,\;
\Phi = \Phi + \sqrt{2} \th \psi_{\Phi} + \th \th F_{\Phi}
 \;\;,\;\;
\\
Q_{i}&=& Q_{i} + \sqrt{2} \th \psi_{Q \; i} + \th \th F_{Q \; i}
\;, \;
   T_{ij} = T_{ij} + \sqrt{2} \th \psi_{T \; ij} + \th \th F_{T \; ij}
 \;\;,\;\; \\
{\rm while}\;\;
D_{\a} &=& \frac{\p}{\p \th^{\a}} \;\; , \;\;
\bar{D}_{\dot{\a}} =  -\frac{\p}{\p \thb^{\dot{\a}}}
\\
V &=& - \th \s^m \thb v_m + i \th \th \thb \lb - i \thb \thb \th
\l + \half \th \th \thb \thb D \;\;.
\eeqn
  We represent the antisymmetric tensor superfield $T_{ij}$
  as
\beqn
\label{eq:antisym}
 Y \equiv \left( TF \right)_{i}^{\;j}
 = \left(
         \begin{array}{cc}
         A  &  B \\
         C  &  A^{t}
        \end{array}
    \right)
\label{eq:Y}
\eeqn
  with $B^{t} = -B$, $C^{t}=- C$.  We define $\tilde{Y}$ similarly.

  In terms of components, the action reads, with indices suppressed,
\beqn
S_{vec}
&=&
\frac{1}{g^2} Tr (- \frac{1}{4} v_{m n} v^{m n} -
[\cD_{m}, \Phi]^{\dagger} [ \cD^{m}, \Phi ]
- i \lambda \s^{m}
[ \cD_{m} , \overline{\lambda} ]
- i \overline{\psi}  \overline{\s}^{m}
[\cD_{m} , \psi]
\nonumber \\
& &
\qquad
- i \sqrt{2} [ \l , \psi ] \Phi^{\dagger} - i \sqrt{2}
[ \overline{\lambda}   , \overline{\psi} ] \Phi)
\nonumber \\
& &
+ \frac{1}{g^2} Tr \left(
\frac{1}{2} D D  - D [\Phi^{\dagger} , \Phi] + F_{\Phi}^{\dagger} F_{\Phi}
\right)
\eeqn

\beqn
S_{asym} &=&
\frac{1}{g^2}
\left\{
- ( \cD_{m} T )^{*} (\cD^{m} T )
- i {\overline{\psi}}_{T}  \overline{\s}^{m}
\cD_{m} \psi_{T}
- i \sqrt{2} T^{*} \l^{(asym)} \psi_{T}
+ i \sqrt{2} {\overline{\psi}}_{T}{ \overline{\lambda}}^{(asym)}  T
\right.
\nonumber \\
& &
- ( \cD_{m} {\tilde{T}} ) (\cD^{m} {\tilde{T}} )^{*}- i
 {\overline{\psi}}_{{\tilde{T}}}   \overline{\s}^{m}
\cD_{m} \psi_{{\tilde{T}}}
- i \sqrt{2} {\tilde{T}}^{*}  \l^{(asym)} \psi_{{\tilde{T}}}
+ i \sqrt{2} {\overline{\psi}}_{{\tilde{T}}}
{ \overline{\lambda}}^{(asym)}  {\tilde{T}}
\nonumber \\
& &
-2 (\Phi^{*}_{(asym)} T^{*}) (\Phi_{(asym)} T)
-2 ( \tilde{T} \Phi_{(asym)}) (\tilde{T}^{*} \Phi^{*}_{(asym)} )\nonumber \\
& &
- \sqrt{2} ( \psi_{{\tilde{T}}} {\psi}^{(asym)} T +  \tilde{T} {\psi}^{(asym)}
 \psi_T		+ \psi_{{\tilde{T}}} \Phi_{(asym)} \psi_T)
\nonumber\\
& &
- \sqrt{2} ( \overline{\psi}_{T} {\overline{\psi}}^{(asym)} \tilde{T}^{*}
		+  T^{*} {\overline{\psi}}^{(asym)} 
\overline{\psi}_{\tilde{T}}		+ \psi_{T} \Phi_{(asym)}^{*} 
\overline{\psi}_{\tilde{T}} )
\nonumber \\
& &
\left.
+ \sqrt{2} \tilde{T} F_{\Phi}^{(asym)} T
+ \sqrt{2} \tilde{T}^{*} F_{\Phi}^{* (asym)} T^{*}
+ \tilde{T} D^{(asym)} T
+ \tilde{T}^{*} D^{(asym)} T^{*}
\right\}
\eeqn


\beqn
S_{fund}
& = &
+ \frac{1}{g^2} \sum_{f=1}^{n_f}
 (
- ( \cD_{m} Q_{(f)} )^{*} (\cD^{m} Q_{(f)} )
- i {\overline{\psi}}_{Q (f)}  \overline{\s}^{m}
\cD_{m} \psi_{Q (f)}
+ i \sqrt{2} Q_{(f)}^{*} \l \psi_{Q (f)}
- i \sqrt{2} {\overline{\psi}}_{Q (f)}  \overline{\lambda} Q_{(f)}
)
\nonumber \\
& &
+ \frac{1}{g^2} \sum_{f=1}^ {n_f}(
- ( \cD_{m} {\tilde{Q}}_{(f)} ) (\cD^{m} {\tilde{Q}}_{(f)} )^{*}- 
i {\overline{\psi}}_{{\tilde{Q}} (f)}  \overline{\s}^{m}
\cD_{m} \psi_{{\tilde{Q}} (f)}
- i \sqrt{2} \tilde{Q}_{(f)}  \l \psi_{{\tilde{Q}} (f)}
+ i \sqrt{2} {\overline{\psi}}_{{\tilde{Q}} (f)}
  \overline{\lambda} {\tilde{Q}}_{(f)}^{*}
)
\nonumber \\
& & + \frac{1}{g^2}  \sum_{f=1}^{n_f} (
Q^{*}_{(f)} D Q_{(f)}+{\tilde{Q}}_{(f)} D {\tilde{Q}}^{*}_{(f)} )
\nonumber \\
& & +
\frac{1}{g^2}  \sum_{f=1}^{n_f} \{
 - (m_{(f)})^2 ( Q^{*}_{(f)} Q_{(f)}+{\tilde{Q}}_{(f)} {\tilde{Q}}^{*}_{(f)})
-  m_{(f)} ( {\tilde{\psi}}_{Q {(f)}} \psi_{Q {(f)}}
+ \bar{{\tilde{\psi}}}_{Q {(f)}} \psib_{Q {(f)}} )
\nonumber\\
& &
- \sqrt{2}
(
Q^{*}_{(f)} \Phi^{\dagger} Q_{(f)} + \tilde{Q}_{(f)} \Phi^{\dagger}
 \tilde{Q}^{*}
+ Q_{(f)}^{*} \Phi Q_{(f)} + \tilde{Q}_{(f)} \Phi \tilde{Q}_{(f)}^{*}
)
\nonumber\\
& &
-2 Q_{(f)}^{*} \Phi^{\dagger} \Phi Q_{(f)} -2 \tilde{Q}_{(f)}
 \Phi^{\dagger} \Phi \tilde{Q}_{(f)}^{*}
\nonumber\\
& &
- \sqrt{2} ( \psi_{{\tilde{Q}}(f)} {\psi} Q_{(f)}	
	+  \tilde{Q}_{(f)}  {\psi} \psi_{Q (f)}	
	+ \psi_{{\tilde{Q}}(f)} \Phi \psi_{Q (f)})
\nonumber\\
& &
- \sqrt{2} ( \overline{\psi}_{Q (f)} {\overline{\psi}} \tilde{Q}_{(f)} ^{*}
		+  Q_{(f)} ^{*} {\overline{\psi}} 
\overline{\psi}_{\tilde{Q} (f)}		+ 
\psi_{Q (f)} \Phi^{\dagger} \overline{\psi}_{\tilde{Q} (f)} )
\nonumber\\
& &
+ \sqrt{2} \tilde{Q}_{(f)}  F_{\Phi} Q_{(f)}+
 \sqrt{2} \tilde{Q}_{(f)} ^{*} F_{\Phi}^{\dagger} Q^{*}_{(f)}
\}
\eeqn

where
\beqn
D_i^{\;\;\; j} &=& [\Phi^{\dagger} , \Phi]_i^{\;\;\; j}
	+ \sum_{f=1}^{n_f} 	(
	Q^{* \; j}_{(f)}  Q_{(f) \; i}
	+{\tilde{Q}}^j_{(f)}  {\tilde{Q}}^{*}_{(f) \; i}
	)
	+2 T^{* \; jk} T_{ki} + 2 \tilde{T}^{jk} \tilde{T}^{*}_{ki}
\label{eqn;D}
\\
F_{\Phi \; i}^{\;\;\;\;\;\; j} &=&
	- \sum_{f=1}^{n_f} 	(
	\sqrt{2} Q^{* \; j}_{(f)}  \tilde{Q}^{*}_{(f) \; i}	)
	- \sqrt{2} T^{* \; jk} T^{*}_{ki}
\label{eqn;FPhi}
\eeqn

  Here ${\cal D}_{m} = i v_{m}$ with $v_{m}$ in  appropriate representations.
 $\Phi_{(anti)}$ , $\psi^{(anti)}$ and $F_{\Phi}^{(anti)}$ are  the fields
 in anti-symmetric representation.

  As is discussed in introduction, the model after  the four dimensional
 compactification possesses  the  exact
 quantum  moduli space which  describes in $F$ theory on $K3$  the local
 deformation of four seven branes away from an orientifold surface.
  For this to hold,  we have to set $n_{f}=4$ and keep  nonvanishing mass
 parameters.  In a more generic situation,   only the global cancellation  of
  the charge associated with eight form potential  is required and
 in this case $n_{f}=16$.
  It remains to be seen whether our model is able to provide a constructive
 framework to this general situation.

%
%

\section{Dynamical and Kinematical Supercharges}

  Let us see how many superchages our model possesses.
It is straightforward to check that the action of our $USp(2k)$ matrix model
 is invariant under the dynamical supersymmetry transformations:
\beqn
\d^{(1)} v_m &=& -i \xib \sb_m \l + i \lb \sb_m \xi -i \etab \sb_m \psi + i
\psib \sb_m \eta
\nonumber \\
\d^{(1)} \l &=& \s^{mn} \xi v_{mn} + i \xi D  - i \sqrt{2}
\s^m \etab \cD_m \Phi
- \sqrt{2} \eta F_{\Phi}
\nonumber \\
\d^{(1)} \Phi &=& \sqrt{2} \xi \psi - \sqrt{2} \eta \l
\nonumber \\
\d^{(1)} \psi &=&  i \sqrt{2} \s^m \xib \cD_m \Phi + \s^{mn}
\eta v_{mn} + i \eta D
+ \sqrt{2} \xi F_{\Phi}
\nonumber\\
\d^{(1)} T &=& \sqrt{2} \xi \psi_T - \sqrt{2} \etab \psib_{\tilde{T}}
\nonumber \\
\d^{(1)} {\tilde{T}}^{*} &=& \sqrt{2} \xib \psib_{\tilde{T}}
                        + \sqrt{2} \eta \psi_T
\nonumber \\
\d^{(1)} \psi_T  &=&  + i \sqrt{2} \s^m \xib \cD_m T
+ i \sqrt{2} \s^m \etab \cD_m {\tilde{T}}^{*}
+ \sqrt{2} \xi F_{T}
+ \sqrt{2} \eta F_{T (T \rightarrow \tilde{T}^{*} ,
 \tilde{T}^{*} \rightarrow -T)}
\nonumber \\
\d^{(1)} \psib_{\tilde{T}}  &=&
- i \sqrt{2} \xi \s^m  \cD_m {\tilde{T}}^{*}
+ i \sqrt{2} \eta \s^m  \cD_m T
+ \sqrt{2} \xib F^{*}_{\tilde{T}}
+ \sqrt{2} \etab F^{*}_{\tilde{T} (T \rightarrow \tilde{T}^{*} ,
 \tilde{T}^{\dagger} \rightarrow -T)}
\nonumber\\
\d^{(1)} Q &=& \sqrt{2} \xi \psi_Q - \sqrt{2} \etab \psib_{\tilde{Q}}
\nonumber \\
\d^{(1)} {\tilde{Q}}^{*} &=& \sqrt{2} \xib \psib_{\tilde{Q}}
                        + \sqrt{2} \eta \psi_Q
\nonumber \\
\d^{(1)} \psi_Q  &=&
+ i \sqrt{2} \s^m \xib \cD_m Q
+ i \sqrt{2} \s^m \etab \cD_m {\tilde{Q}}^{*}
+ \sqrt{2} \xi F_{Q}
+ \sqrt{2} \eta F_{Q (Q \rightarrow \tilde{Q}^{*} ,
 \tilde{Q}^{*} \rightarrow -Q)}
\nonumber \\
\d^{(1)} \psib_{\tilde{Q}}  &=&
- i \sqrt{2} \xi \s^m  \cD_m {\tilde{Q}}^{*}
+ i \sqrt{2} \eta \s^m  \cD_m Q
+ \sqrt{2} \xib F^{*}_{\tilde{Q}}
+ \sqrt{2} \etab F^{*}_{\tilde{Q} (Q \rightarrow \tilde{Q}^{*} ,
 \tilde{Q}^{*} \rightarrow -Q)} \;\;,
\nonumber \\
\;\;
\eeqn
where
$D$ and $F_{\Phi} $ are given by (\ref{eqn;D}) (\ref{eqn;FPhi}), and
\beqn
F_{T \; ij} = - \sqrt{2} \left( \Phi^{*}_{(asym)}
 \right)_{ij}^{\;\;\;k \ell}
\tilde{T}^{*}_{k \ell}
\;\; , \;\;
F_{Q \; i} = - m \tilde{Q}^{*}_{i}
	- \sqrt{2} \tilde{Q}^{*}_{k} \Phi^{* k}_{ \;\;\;\; i} \;\; . \;\;
\eeqn
\beqn
F^{*}_{\tilde{T} \; ij} = - \sqrt{2} \left( \Phi_{(asym)}
 \right)_{ij}^{\;\;\;k \ell}
\tilde{T}_{k \ell}
\;\; , \;\;
F^{*}_{Q \; i} = - m \tilde{Q}_{i}
	- \sqrt{2}  \Phi^{ \;\;\; j}_{ i} \tilde{Q}_{j} \;\; . \;\;
\eeqn

The kinematical supersymmetry transformations are
\beqn
& &
\d^{(2)} v_m = 0 \;\; , \;\;
\d^{(2)} \l = 0 \;\; , \;\;
\d^{(2)} \Phi = 0 \;\; , \;\;
\d^{(2)} \psi = 0 \;\; , \;\;
\nonumber\\
& &
\d^{(2)} Q = 0 \;\; , \;\;
\d^{(2)} {\tilde{Q}}^{*} = 0 \;\; , \;\;
\d^{(2)} \psi_Q  = 0 \;\; , \;\;
\d^{(2)} \psib_{\tilde{Q}} = 0 \;\; , \;\;
\nonumber\\
& &
\d^{(2)} T = 0 \;\; , \;\;
\d^{(2)} {\tilde{T}}^{*} = 0 \;\; , \;\;
\d^{(2)} \psi_T  = \zeta \;\; , \;\;
\d^{(2)} \psib_{\tilde{T}}  = \bar{\tilde{\zeta}} \;\; .
\eeqn
So our model has eight dynamical supercharges and eight kinematical ones.
This is the proper number of supercharges in order for this model to be
 interpretable as critical string.
Up to field dependent gauge transformations and equations of motion for
 the fermionic fields, we obtain the following commutation relations.
\beqn
& & [ \d^{(1)}_{\xi , \eta} , \d^{(1)}_{\xi' , \eta'} ] T  =  0
\nonumber \\
& & [ \d^{(1)}_{\xi , \eta} , \d^{(1)}_{\xi' , \eta'} ] \psi_{T}  =  0
\nonumber \\
& & [ \d^{(1)}_{\xi , \eta} , \d^{(1)}_{\xi' , \eta'} ] {\tilde{T}}^{*}  =  0
\nonumber \\
& & [ \d^{(1)}_{\xi , \eta} , \d^{(1)}_{\xi' , \eta'} ] \psib_{\tilde{T}}
  =  0 \;\;.
\eeqn
We also have the following commutation relations;
\beqn
& & [ \d^{(1)}_{\xi , \eta} , \d^{(2)}_{\zeta , \tilde{\zeta}} ] T
= \sqrt{2} ( \xi \zeta - \bar{\eta }  \bar{ \tilde{\zeta}}  )
\nonumber\\
& &[ \d^{(1)}_{\xi , \eta} , \d^{(2)}_{ \zeta , \tilde{\zeta} } ] \psi_{T}
 =  0
\nonumber \\
& & [ \d^{(1)}_{\xi , \eta} , \d^{(2)}_{\zeta , \tilde{\zeta}} ]
 {\tilde{T}}^{*}
=
\sqrt{2} ( \xi \zeta + \bar{\eta }  \bar{ \tilde{\zeta}}  )
\nonumber \\
& & [ \d^{(1)}_{\xi, \eta} , \d^{(2)}_{\zeta , \tilde{\zeta}} ]
 \psib_{\tilde{T}}  =  0
\eeqn
\beqn
& & [ \d^{(2)}_{\zeta , \tilde{\zeta}} , \d^{(2)}_{\zeta' , \tilde{\zeta}'} ]
 T  =  0
\nonumber \\
& & [ \d^{(2)}_{\zeta , \tilde{\zeta}} , \d^{(2)}_{\zeta' , \tilde{\zeta}'} ]
 \psi_{T}  =  0
\nonumber \\
& & [ \d^{(2)}_{\zeta , \tilde{\zeta}} , \d^{(2)}_{\zeta' , \tilde{\zeta}'} ]
 {\tilde{T}}^{*}  =  0
\nonumber \\
& & [ \d^{(2)}_{\zeta , \tilde{\zeta}} , \d^{(2)}_{\zeta' , \tilde{\zeta}'} ]
 \psib_{\tilde{T}}  =  0
\eeqn
   The combination $\d^{(1)} \pm \d^{(2)}$, therefore,
  forms  supersymmetry algebra of sixteen supercharges which closes into
 translation of the four of the bosonic matrices in the antisymmetric
 representation.

%
%
%
%
\section{Vacuum configuration}

 Let us find a configuration having vanishing action, which is a
 particular classical solution of the model.
  This tells us how many spacetime coordinates are generated from our model.
   We set all fermions zero in the action. We first demand
\beqn
\label{eq:demand}
  v_{mn} &=& 0 \nonumber \\
  \left[ {\cal D}_{m}, \Phi \right]  &=&  i \left[ v_{m} , \Phi \right]= 0
 \nonumber \\
 {\cal D}_{m} Q_{f} &=& {\cal D}_{m} \tilde{Q}_{f} =0 \;\;.
\eeqn
  We see that  all of $v_{m}, \Phi$, and $\Phi^{\dagger}$ lie on the
 Cartan subalgebra of $usp(2k)$, namely
\beqn
\label{eq:Md}
 N=0\;\; {\rm and}\;\; M = d = {\rm  diagonal}\;\;
\eeqn
 in $X$ of eq.~(\ref{eq:usp}).  In addition,
\beqn
\label{eq:MQf}
 Q_{f}= \tilde{Q}_{f} =0 \;\;.
\eeqn
   As for the antisymmetric tensor fields, we first examine
\beqn
\mid \mid {\cal D}^{m}\left( asym \right)_{ij}^{\;k \ell} T_{k \ell}
 \mid \mid^{2} \;\;,
\eeqn
  where ${\cal D}^{m}\left( asym \right) = i v^{m\;(r)} t^{(r)}(asym)$.
  This is expressible as
\beqn
 tr  \left[ {\cal D}_{m}, Y \right]  \left[ {\cal D}_{m},
 Y^{\dagger} \right] \;\;.
\eeqn
  In general, the commutator  $\left[ X, Y \right]$ with $X \in usp(2k)$ is
 written in terms of $k \times k$ blocks as
\beq
[X,Y] =
\left(
\begin{array}{cc}
[M,A] - (-NC + BN^{*})            &  MB - (MB)^t - AN + (AN)^t \\
N^{*}A - (N^{*}A)^t -CM + (C M)^t  &  [M,A]^t - (-NC+ BN^{*})^t
\end{array}  \;\;
\right) \;\;.
\eeq
  When $X$ is restricted to the Cartan subalgebras,
 the condition  that the commutator vanish
 $ \left[ X, Y \right] =0 $ implies
\beqn
 \left[ d ,A \right] = 0 \;, \;\; \left( d_{i}+ d_{j}
 \right) B_{ij}=0 \;\; {\rm and}  \;\;  \left( d_{i}+ d_{j}
 \right) C_{ij}=0       \;\;  {\rm not~ summed}
\;\;,
\eeqn
and therefore
\beqn
\label{eq:AB}
  A &=&   a = {\rm diagonal}\;\;, \;\; B = C = 0 \;\; \nonumber \\
  \tilde{A} &=& \tilde{a} = {\rm diagonal}\;\;, \;\; \tilde{B} =
  \tilde{C}=  0 \;\;
\eeqn
in eq.~(\ref{eq:Y}).

  Under eqs.~(\ref{eq:demand}),(\ref{eq:MQf}),  with all fermions set  zero,
  the remaining part of the action  $S_{res}$  is
\beqn
\label{eq:VDmat}
S_{res}&=&
\frac{1}{g^2} Tr \left\{
\frac{1}{2} D D  - D [\Phi^{\dagger} , \Phi] + F_{\Phi}^{\dagger} F_{\Phi}
\right\}
\nonumber \\
& &
\;\;\;\;\;\;\;\;\;\;
-2 (\Phi^{*}_{(asym)} T^{*}) (\Phi_{(asym)} T)
-2 ( \tilde{T} \Phi_{(asym)}) (\tilde{T}^{*} \Phi^{*}_{(asym)} )\nonumber \\
& &
\;\;\;\;\;\;\;\;\;\;
+ \sqrt{2} \tilde{T} F_{\Phi}^{(asym)} T
+ \sqrt{2} \tilde{T}^{*} F_{\Phi}^{* (asym)} T^{*}
+ \tilde{T} D^{(asym)} T
+ \tilde{T}^{*} D^{(asym)} T^{*}
\nonumber \\
&=&
\frac{1}{g^2} Tr \left\{
- \frac{1}{2} ([\Phi^{\dagger} , \Phi]
		+ [Y^{\dagger} , Y] + [\tilde{Y}^{\dagger} , \tilde{Y} ])^2
\right.
\nonumber \\
& &
\;\;\;\;\;\;\;\;\;\;
\left.
- 2 ( [\tilde{Y}^{\dagger} , Y^{\dagger} ] [Y, \tilde{Y}]
	+[\tilde{Y} ,\Phi]  [\Phi^{\dagger} ,\tilde{Y}^{\dagger}]
	+[ Y^{\dagger} ,\Phi^{\dagger} ] [\Phi , Y]
)
\right\}
\eeqn
Eq.~(\ref{eq:VDmat}) vanishes for the configuration  satisfying
  eqs.~(\ref{eq:Md}) and (\ref{eq:AB}).

  We conclude that the vacuum configuration is represented by
\beq
v_M =
\left(
\begin{array}{cccccc}
p_M^1 &      & & &  \\
      & \ddots  & & &  \\
      &         & p_M^k & &  \\
      &         &       & sgn(M) p_M^{1} &  \\
      &         &       &                  & \ddots &  \\
      &         &       &                  &        & sgn(M) p_M^{k} 
\end{array}
\right)
 \equiv  {\rm diag}\; v_{M}^{(class)} \;,
\eeq
and
\beqn
 Q_{f} = \tilde{Q}_{f} = Q^{*}_{f} = \tilde{Q}^{*}_{f} =0\;\;.\\
sgn(M) =
\left\{
\begin{array}{cc}
 -1 & M= 0, \cdots ,5 \\
+1  & M = 6 , \cdots ,9
\end{array}
\right. \;\;,
\eeqn
where
\beq
\label{eq:vm}
v_M \equiv \left( v_{m} , \frac{ \Phi + \Phi^{\dagger}}{\sqrt{2}}
,\frac{ \Phi - \Phi^{\dagger}}{\sqrt{2}i}
, \frac{ Y + Y^{\dagger}}{\sqrt{2}}
,\frac{ Y - Y^{\dagger}}{\sqrt{2}i}
,\frac{ \tilde{Y} + \tilde{Y}^{\dagger}}{\sqrt{2}}
,\frac{ \tilde{Y} - \tilde{Y}^{\dagger}}{\sqrt{2}i}
\right)  \;\;\;.
\eeq

  The spacetime coordinates generated extend not only to the six directions
 obtained from the gauge fields and the  adjoint scalars lying on the
 Cartan subalgebra of the $usp(2k)$ but also
  to the four additional directions  from the antisymmetric tensor fields.
 
  It is relatively clear  that the one-loop stability of this vacuum
  is ensured  by supersymmetry.  We consider the second order fluctuation
  from ${\rm diag}\;v_{M}^{(class) }$  and compute determinants.
  Let  the adjoint action
   $\hat{P}_{M}$ on  matrix $X$ be $\hat{P}_{M} X = \left[ {\rm diag}\;
 v_{M}^{(class)}, X \right]$.  Following  \cite{KEK}, the one-loop effective
  action obtained from the bosonic, fermionic and ghost degrees of freedom
  is  $ \left(\frac{1}{2} \cdot 10 - \frac{1}{4} \cdot16 -1 \right) Tr \log
 \left( \hat{P}^{M} \hat{P}_{M} \right)$ and vanishes.

 We  would now like to have a more definite physical interpretation of
 this model than  those discussed briefly in section $1$ and section $2$.
 At the same time  we would like to
 clarify the issue of  the number of spacetime dimensions. Let us 
 recall that ten of the noncommuting coordinates $v_{M}$
  ( eqs.~(\ref{eq:vm}), (\ref{eq:usp}), (\ref{eq:antisym})),  which are
 dynamical variables,  satisfy
\beqn
\label{eq:rel}
v_{i}^{t} &=& - F v_{i} F^{-1}\;\;    i= 0 \sim 5\;\;, \nonumber \\
v_{I}^{t} &=& \; F v_{I} F^{-1} \;\; I = 6 \sim 9 \;\;.  
\eeqn
 The $v_{M}$'s are noncommuting analog of the ten string coordinates
   $X_{M}$ in the standard first quantized approach.  The operation  $F$
  is matrix analog of the twist operation $\Omega$.
The classical counterpart of eq.~(\ref{eq:rel})  is  therefore
\beqn
\label{eq:clrel}
X_{i} &=& - \Omega X_{i} \Omega^{-1} \;\; i= 0 \sim 5    \nonumber \\
X_{I} &=& \;  \Omega X_{I} \Omega^{-1} \;\; I = 6 \sim 9 \;\;.  
\eeqn
 The presence of a four dimensional fixed surface ( orientifold surface)
 becomes clear  from this equation (\ref{eq:clrel}).

Eq.~(\ref{eq:rel}) is also equations of embedding  $v_{M}$'s into $U(2k)$
 matrices.   In fact, via this embedding, the part of the action 
  which do not invlove fundamentals is
 obtained from zero-dimensional reduction of ten-dimensional  super
 Yang-Mills theory by the projection.
 This dimensionally reduced model has been interpreted as
 a matrix model of type $IIB$ superstring \cite{KEK}.
 We conclude  that  our model not only  provides an exact
  $F$ theory compactification through  the procedure of \cite{WT}
   applied to  $v_{m}$'s ($m =0,1,2,3$)  but also is a matrix model
   representing type $IIB$ superstring on $T^{6}/Z^{2}$  orientifold with
 large volume limit taken.
  
  Let us now impose periodicities on infinite  size matrices
 $v_{m}$'s  ($m =0,1,2,3$)  for all four directions.
   For this, we decompose  $v_{m}$'s into  blocks of $n \times n$  matrices.
  We specify each individual block by a row vector $\vec{a} = (a_{1}, \cdots
 a_{4})$ and a column vector $\vec{b} = (b_{1}, \cdots b_{4})$:
 $\left( v_{m}\right)_{\vec{a}, \vec{b}} \equiv  \sqrt{\alpha^{\prime}}
 \langle \vec{a} \mid  \hat{v}_{m} \mid \vec{b} \rangle $.
  Let the shift vector be
\beqn
 \left( U(i)\right)_{\vec{a}, \vec{b}} =  \left( \prod_{j (\neq i)}
 \delta_{a_{j}, b_{j}} \right) \delta_{a_{i}, b_{i}+1} \;\;.
\eeqn
 The condition to be imposed is
\beqn
\label{eq:period}
U(i) v_{m} U(i)^{-1} &=&  v_{m} - \delta_{m,i} R /\sqrt{\alpha^{\prime}} \;\;.
\eeqn
 The solution  in Fourier transformed space is
\beqn
\langle  \vec{x} \mid \hat{v}_{m} \mid \vec{x}^{\prime} \rangle
 &=& -i \left( \frac{\partial}{\partial x^{m}} +i \tilde{v}_{m}(\vec{x})
 \right)
  \delta^{(4)} \left( \vec{x}- \vec{x}^{\prime} \right)
 \;\;, \\
\tilde{v}_{m}(\vec{x}) &=&  \sum_{\vec{\ell} \in {\cal Z}^{4}}
 \tilde{\tilde{v}}_{m}(\vec{\ell})
  \exp \left( \frac{-i \vec{\ell} \cdot \vec{x}}{\tilde{R}} \right)
 \;\;, \nonumber \\
 \tilde{R} &\equiv& \alpha^{\prime} /R \;\;\;.
\eeqn
The Fourier transform acts as the $T$ dual transformation: it interchanges the
  radius parameter $R$  setting the period of the original matrix index
 with the dual radius $\tilde{R}$  which is the period of the space
  Fourier conjugate to the matrix index.
  The point is that we go to the $R \rightarrow 0$ limit after the Fourier
 transform.  The resulting description of the model in the dual coordinates
  $v_{m}$'s $m =0,1,2,3$ is  the large $k$ limit of the
   ${\cal N}=2$  supersymmetric $USp(2k)$ gauge theory with an antisymmetric
  and $n_{f}$ fundamental hypermultiplets.  The ten noncommuting coordinates
 have classical counterparts which are 
\beqn
\label{eq:clrel2}
X_{i} &=& - \Omega X_{i} \Omega^{-1} \;\;\; i= 4, 5    \nonumber \\
\tilde{X}_{\ell} &=& \;\; \Omega \tilde{X}_{\ell} \Omega^{-1} \;\;\;
  \ell = 0,1,2,3   \nonumber \\
X_{I} &=&  \;\; \Omega X_{I} \Omega^{-1} \;\;\; I = 6 \sim 9 \;\;.   
\eeqn 
  Here $\tilde{X}_{\ell} \equiv \tilde{X}_{\ell\;R} - \tilde{X}_{\ell\;L}$
  denote the dual coordinates of
 $X_{\ell} \equiv \tilde{X}_{\ell\;R}+ \tilde{X}_{\ell\;L}$.
 The fixed orientifold surface is now eight dimensional as minus sign
 appears   twice ($i=4,5$) in eq.~(\ref{eq:clrel2}). The model in this limit
   therefore represents the nonperturbative completion of type $IIB$
 on large volume $T^{2}/Z^{2}$ orientifold, namely on $CP^{1}$.

  As it stands, the model produces only ten spacetime dimensions as
 eigenvalue distributions of the ten matrix coordinates. 
 The original argument of  ref.~\cite{V} that $F$ theory  is 
   theory in twelve spacetime dimensions  becomes in the present context
 of our matrix model representing $F$ theory as follows.
  The low energy effective action of the four dimensional
   $USp(2k)$  ${\cal N}=2$ susy gauge theory above,
     which is our matrix model in the compactification described,
    has been exactly  determined \cite{SW}.  The effective
 running coupling  $\tau$  depends upon $tr \Phi^{2}/2 \equiv u$ :
  $\tau = \tau(u)$.  The work of \cite{SW} exactly determines this function
 and  this is precisely  the moduli of the torus which  is the fiber of the
 elliptic fibered $K3$ surface  with $CP^{1}$  the base labelled by $u$.
  This is the only rationale for regarding this torus to represent two
 additional spacetime.
  Coordinates which parametrize this torus do not, however, manifest
   themselves in the present framework: only its moduli appear. 
  Even if one is willing to take this twelve dimensional viewpoint,
 the two additional  dimensions are treated  very differently from the
 remaining ten dimensions which  are directly related to the noncommuting
 coordinates as dynamical variables \footnote{We should, however, note that
 the possibility of twelve spacetime dimensions is suggested
 in ref. \cite{12D} as well as in the recent work based on current
 algebra \cite{Suga} and in the one on topological matrix model \cite{HK}.}


\section{Acknowledgements}
  We thank Shinji Hirano and Asato Tsuchiya for helpful discussion
 on this subject.



\newpage

\begin{thebibliography}{99}

\bibitem{Sen}
A. Sen,
{\sl Nucl.Phys.}{\bf B475}(1996)562.

\bibitem{Witten}
 E. Witten, {\sl Nucl. Phys.} {\bf B460} (1996)335.

\bibitem{BFSS}
T. Banks, W. Fischler, S.H. Shenker, L. Susskind, 
{\sl Phys. Rev.} {\bf D55} (1997)5112.

\bibitem{M}
 E. Witten, {\sl Nucl. Phys.} {\bf B443} (1995)85.

\bibitem{KEK}
N. Ishibashi, H. Kawai, Y. Kitazawa and A. Tsuchiya,
 {\sl Nucl. Phys.} {\bf B498} (1997)467.
\\
M. Fukuma, H. Kawai, Y. Kitazawa and A. Tsuchiya, preprint hep-th/9705128.

\bibitem{2B}
A. Fayyazuddin, Y. Makeenko, P. Olsen, D. J. Smith and K. Zarembo,
preprint hep-th/9703038;
\\
I. Chepelev and A. A. Tseytlin, preprint hep-th/9705120

\bibitem{comp}
  Partial list includes 
L. Motl, preprint hep-th/9612198;
N. Kim and S.J. Rey, preprint hep-th/9701139;
R. Dijkgraaf, E. Verlinde and H. Verlinde, preprint hep-th/9703030;
A. Fayyazuddin and D.J. Smith, preprint hep-th/9703208;
T. Banks and L. Motl, preprint hep-th/9703218;
P. Horava, preprint hep-th/9705055;
N. Kim and S.J. Rey, preprint hep-th/9705132;
S. Sethi and L. Susskind, preprint hep-th/9702101;
T. Banks and N. Seiberg, preprint hep-th/9702187;

\bibitem{het}
  For the heterotic case,
N. Kim and S.J. Rey, preprint hep-th/9701139;
T. Banks and L. Motl, preprint hep-th/9703218;
D.A. Lowe, preprint hep-th/9704041;
S.J. Rey, preprint hep-th/9704158;
P. Horava, preprint hep-th/9705055.

\bibitem{WT}
W. Taylor IV , {\sl Phys. Lett.} {\bf B394} (1997)283;
O.J. Ganor, S. Ramgoolam and W. Taylor IV,
{\sl Nucl.Phys.}{\bf B492}(1997)191.


\bibitem{V}
C. Vafa,
{\sl Nucl.Phys.}{\bf B469}(1996)403.

\bibitem{PW}
 J. Polchinski and E. Witten, {\sl Nucl. Phys.} {\bf B460} (1996)525.

\bibitem{AIKT}
 Y. Arakane, H. Itoyama, H. Kunitomo and A. Tokura, 
{\sl Nucl. Phys.} {\bf B486} (1997)149.

\bibitem{SW}
 N. Seiberg and E. Witten, 
{\sl Nucl. Phys.} {\bf B426} (1994)19 ,
{\sl Nucl. Phys.} {\bf B431} (1994)484.

\bibitem{BDS}
T. Banks, M.R. Douglas, N. Seiberg,
{\sl Phys.Lett.}{\bf B387}(1996)278.
\bibitem{DLSASYT}
M.R. Douglas, D.A. Lowe, J.H. Schwarz, 
{\sl Phys.Lett.}{\bf B394}(1997)297.
\\
O. Aharony, J. Sonnenschein, S. Yankielowicz and S. Theisen, 
{\sl Nucl.Phys.} {\bf B493} (1997) 177 .


\bibitem{12D}
V. Periwal, preprint hep-th/9611103;
\\
A. A. Tseytlin, preprint hep-th/9612164.

\bibitem{Suga}
H. Sugawara,  preprint hep-th/9708029.

\bibitem{HK}
S. Hirano and M. Kato, preprint hep-th/9708039.

\end{thebibliography}
\end{document}